\documentstyle[aps,epsf,preprint]{revtex}
\begin{document}

\baselineskip=24 pt

%\twocolumn[\hsize\textwidth\columnwidth\hsize\csname@twocolumnfalse\endcsname

\title
{Emergence of hexatic and long-range herringbone order in
two-dimensional smectic liquid crystals : A Monte Carlo study }
\author
{Farhad Shahbazi $^a${\footnote{Electronic address:
shahbazi@cc.iut.ac.ir }} ,Rasool Ghanbari $^b$   }
\address
{\it $^a$ Dept. of Physics , Isfahan University of Technology,
84156, Isfahan, Iran. \\
\it $^b$ Dept. of Physics , Islamic Azad University, Majlesi
branch, 86315/111, Isfahan ,Iran.
\\}

\maketitle

\begin{abstract}
Using a high resolution  Monte Carlo simulation technique based on
multi-histogram method and cluster-algorithm, we have
investigated critical properties of a coupled XY model, consists
of a six-fold symmetric hexatic and a three-fold symmetric
herringbone field, in two dimensions. The simulation results
demonstrate a series of novel continues transitions, in which both
long-range hexatic and herringbone  orderings are established
simultaneously. It is found that the specific-heat anomaly
exponents for some regions in coupling constants space are in
excellent agreement with the experimentally measured exponents
extracted from heat-capacity data near the smecticA-hexaticB
transition
of two-layer free standing films.\\
%\vspace{1mm}
 PACS numbers: 61.30.-v, 64.70.Md
\end{abstract}
\hspace{.3in}
\newpage

\section{Introduction}
The mysterious  critical properties of bulk and thin film liquid
crystals at phase transition between smecteicA (SmA) phase with
liquid like in-plane behaviour  and  HexaticB(HexB) phase with
long-range bond orientational order but short range in-plane
positional order,  has been remained as a challenge for
experimental and theoretical physicists after about 25 years.

The concept of hexatic phase was first introduced in two
dimensional melting theory by Kosterlitz, Thouless, Halperin,
Nelson and Young (KTHNY) {\cite{KT},\cite{hn},\cite{y}}.
According to this theory,  two dimensional systems during melting
transition from solid to isotropic liquid go through an
intermediate phase called hexatic phase for systems that have
six-fold(hexagonal) symmetry in their crystalline ground state.
This hexatic phase displays short range positional order, but
quasi long range bond-orientational order, which is different
from the true long range bond-orientational and quasi long range
positional order in 2D solid phases. It is known that for two
dimensional systems, the transition from the isotropic liquid  to
hexatic phase could be either a KT transition or a first order
transition {\cite{5}}.

The idea of hexatic phase was first applied to three dimensional
systems by Birgeneau and Lister, who showed that some
experimentally observed smectic liquid crystal phases ,consisting
of stacked 2D layers, could be physical realization of 3D
hexatics\cite{birg}. Assuming that the weak interaction between
smectic layers could make the quasi long range order of two
dimensional layers truly long ranged, they suggest that the 3D
hexatic phases in highly anisotopic systems, possess  short range
positional and true long range bond-orientational order.

The first signs for the existence of the hexatic phase in three
dimensional systems were observed in x-ray diffraction study of
liquid crystal compound
65OBC(n-alkyl-4-m-alkoxybiphenyl-4-carboxylate,n=6,m=5)\cite{huang,pindak},
where a hexagonal pattern of diffuse spots was found in intensity
of scattered x-rays. In addition to this hexagonal pattern, it was
also found that some broader peaks were appeared in the diffracted
intensity which indicate the onset of another ordering. These
broad peaks are related to packing of molecules according to the
herringbone structure perpendicular to the smectic layer stacking
direction, although the range of herringbone ordering was not
determined by a detailed investigation. The accompanying of the
long range hexatic and herringbone orders make this phase a
physically rich phase which simply is called Hexatic-B (HexB)
phase. When temperature is decreased, the HexB phase transforms
via a first order phase transition into the crystal-E (CryE)
phase, which is a 3D plastic crystal exhibiting  long range
herringbone orientational ordering. Subsequently, it was found
that other components in nmOBC homologous series (like 37OBC and
75OBC) and a number of binary mixtures of
n-alkyl-4'-n-decycloxybiphenyl-4-carboxilate (n(10)OBC) with n
ranging from 1 to 3 and also  represent smA-HexB transition. In
summary the most of $nm$OBC homologous series undergo the
following bulk transition sequences
Isotropic-SmA-HexB-CryE-CryK,where CryK is the rigid crystal
structure, stable at room temperature.

The sixfold symmetry of hexatic phase suggests that
bond-orientational order parameter to be  defined by
$\Psi_{6}=|\Psi_{6}|\exp(i6\psi_{6})$ describing the sixfold
azimuthal modulation. The U(1) symmetry of the $\Psi_{6}$,
implies that SmA-HexB transition be a member of XY universality
class. However, heat capacity measurements on bulk samples of
65OBC \cite{huang,huang2} and other calorimetric studies on many
other components in the nmOBC homologous series
\cite{huang,pitch} have yielded very sharp specific heat
anomalies near SmA-HexB transition with no detectable thermal
hystersis and with very large value for the heat capacity
critical exponent, $\alpha\approx0.6$. These results indicate
that this is a continues (second order) phase transition, but not
belonging to The 3D XY universality class, for which the specific
heat critical exponent is nearly zero
($\alpha\approx-0.007$\cite{zinn}). On the other hand, the other
static critical exponents determined  from thermal conductivity
($\eta=-0.19$)and  birefringence experiments ($\beta=0.19$)
\cite{huang}, all differ from the 3D XY values, indicating a
novel phase transition with probably a new universality class.

These unusual behaviour also occurs in two dimensional liquid
crystal compounds undergoing bulk  hexatic transitions. The heat
capacity measurement studies of (truly two-dimensional) two-layer
free standing films of different nmOBC compound result a second
order SmA-HexB transition, with a diverging specific-heat anomaly
described by  exponent $\alpha=0.31\pm 0.03$\cite{huang,huang3}.
This is obviously  in contrast with the usual broad and
nonsingular specific-heat hump of the KT transition far above
$T_{c}$, predicted in two-dimensional melting theory. On the other
hand, electron diffraction studies on $nm$OBC compound films
revealed weak herringbone orders in hexatic phases, suggesting
that SmA-HexB transition can not be described simply by a unique
XY order parameter and the discrepancy between the experimental
and two dimensional melting theory could be due to the presence
of herringbone order in addition to bond-orientational order in
such compounds.

In the light of these observations,  Bruinsma and
Aeppli{\cite{BA}}  formulated a Ginzburg-Landau theory that
included both hexatic and herringbone order. There are three
inequivalent  orientation for herringbone pattern on a triangular
lattice, so the herringbone order parameter should be three-fold
symmetric. Nevertheless, the broadness of x-ray diffracted peaks
associated to herringbone order made Bruinsma and Aeppli to
consider it short-ranged and associate  an XY order parameter with
two fold symmetry for herringbone ordering
$(\Phi_{2}=|\Phi_{2}|\exp(i2\phi_{2}))$. Based on symmetry
arguments, they also made a minimal coupling between the hexatic
and herringbone order parameters as
$V_{hex-her}=hRe(\Psi_{6}^{*}\Phi_{2}^{3})$. Microscopically, the
origin of this coupling could be the  anisotropy presented in
liquid crystals molecular structures\cite{michel1,michel2}.

In the mean field approach their results indicate that the
SmA-HexB transition should be continuous. However one-loop
renormalization calculations show that short range molecular
herringbone correlations coupled to the hexatic ordering drive
this transition first order, which becomes second order at a
tricritical point{\cite{BA}}. Their result indicates the
existence of two tricritical points, one for the transition
between SmA phase ($\Psi=0, \Phi=0$) and the stacked hexatic
phase ($\Psi\neq 0, \Phi=0$), and another for the transition
between the SmA and the phase possessing both hexatic and
herringbone order ($\Psi\neq 0, \Phi\neq 0$). Therefore, They
concluded that the occurrence of phase transition  near the
tricritical points, with heat capacity exponent $\alpha=0.5$,
would be a good explanation for large heat capacity exponents
observed in the  experiments. Recently, the RG calculation of BA
model has been revised in \cite{kohan} which resulted in finding
another non-trivial fixed point missed in original work of
Bruinsma and Aeppli. But it has been shown that this new fixed
point is unstable in one loop level (order of $\epsilon$), which
refuses this fixed point to represent a novel phase transition.
Indeed, the limitations of  RG methods which mostly rely on
perturbation expansions, make them insufficient for accessing the
strong coupling regimes where one expect that some kind of new
treatment to appear. For this purpose, the numerical simulations
would be useful.

The first  numerical simulations for investigating the nature of
the SmA-HexB transition in 2D systems have been done by Jiang et
al who have used a model consists of a 2D lattice of coupled XY
spins based on the BA Hamiltonian in strong coupling
limit\cite{mc1,mc2}. Their simulation results suggest the
existence of a new type phase transition in which two different
orderings are simultaneously established through a continuous
transition with heat capacity exponent $\alpha=0.36\pm0.05$, in
good agreement with experimental values. Recently, we have carried
out a high-resolution Monte Carlo simulation, based on
multi-histogram method, on BA model in three dimensions. Our
results revealed the existence of a tricritical point on the
transition line between SmA and hexatic+herringbone phases, but
not  any tricritical point on isotropic-Hexatic transition
line{\cite{ghanbari}}.

However, While the occurrence of SmA-HexB in the vicinity of a
tricritical point might be convincing reason for its observed
large heat capacity exponents, some other questions remain
unsolved. One question is that why seven different liquid crystal
compounds nmOBC and five binary mixtures n(10)OBC, with very
different SmA-HexB temperature ranges(which effect the coupling
of two order types) yield approximately the same value
$\alpha\approx0.6$ and should all be in the immediate vicinity of
a particular thermodynamic point. For example, the herringbone
peaks observed in x-ray diffraction studies of 75OBC is weaker
than those of 65OBC, hence if 65OBC is near a tricitical point
then 75OBC should be further removed from this point. Yet the same
specific-heat critical exponents has been obtained for these two
compounds. The other problem concerns the mixture of 3(10)OBC
(3-alkyl-4'-n-decycloxybiphenyl-4-carboxilate) and PHOAB
(4-propionyl-4'-n-heptanoyloxyazo-benzene) compounds with PHOAB
concentration between 30 and 70 percent, for which one expect
very small herringbone fluctuations (because of large HexB
temperature range) and therefore the SmA-HexB transition must be
second order but a first order transition were found for these
mixtures.

Therefore,  this idea that the coupling of the hexatic order with
short-ranged herringbone fluctuations is responsible for the
unexpected critical behaviour of SmA-HexB transitions, has been
faced with serious challenges by the above consideration. To best
of our knowledge no detailed X-ray or electron diffraction
studies have been done to determine the range of herringbone
fluctuations, so it might not be truly short-ranged. On the other
hand, the studies on thin-film heat-capacity data have suggested
the probability of the existence of long-range herringbone order
in a system with long-range bond orientational order and
short-range transational order\cite{huang}.

In views of the above remarks, we was motivated to investigate the
critical properties of a coupled $XY$ model consisting  of a
hexatic order parameter with sixfold symmetry and a three-fold
symmetric herringbone order. For this purpose we employed a
high-resolution Monte Carlo technique to derive the critical
temperatures and the critical exponents of this model over some
ranges of coupling constants.

The rest of this paper is organized as follows. In section. II, we
introduce model Hamiltonian  and give a brief introduction to
Wolf's embedding method for reducing the critical slowing down and
correlation between the measurements, the optimized Monte Carlo
method based on multiple histograms and also Some methods for
analyzing the Monte Carlo data to determine the order of
transitions,  critical temperatures and critical exponents. The
simulation results and discussion is given in section III and
conclusions will appear in section IV.

\section{Monte Carlo simulation}

\subsection{Model}

Recalling  the six-fold symmetry of hexatic order and  two-fold
symmetry of the long range herringbone order, the Hamiltonian
which describes both orderings ought to be invariant with respect
to the transformation $\Phi\rightarrow\Phi+n(2\pi/3)$ and
$\Psi\rightarrow\Psi+m(2\pi/6)$ where $n$ and $m$ are integers.
Thus to lowest order in $\Psi$ and $\Phi$, one can write the
following  Hamiltonian for this model:

\begin{eqnarray}
H=&-&J_{1}\sum_{<ij>} \cos(\Psi_{i}-\Psi_{j})- J_{2}\sum_{<ij>}
\cos(\Phi_{i}-\Phi_{j})\nonumber\\
&-&J_{3}\sum_{i} \cos(\Psi_{i}-2\Phi_{i}),
\end{eqnarray}

where the coefficients $J_1$ and $J_2$ are the nearest-neighbor
coupling constants for the bond-orientational order $(\Psi)$ and
herringbone order $(\Phi)$, respectively. The coefficient $J_3$
denotes the coupling strength between these two types of order at
the same 3D lattice site. we are interested in situations in
which $\Psi$ and $\Phi$ are coupled relatively strongly. Therefore
for the beginning we fixed $J_3=2.0$,larger that both $J_1$ and
$J_2$ for all the simulations. Assuming  $J_{1}$ much larger than
$J_{2} ($ $J_{1}>>J_{2}$), for sufficiently high temperatures,(say
$T>J_{3}$), the system is in completely disordered phase. For
$T_{c1}<T<J_3$, the system remains disordered but the phases of
the two order parameters become coupled through the
herringbone-hexatic coupling term $J_{3}$. because of the XY
symmetry of bond-orientational order ,for $T_{c2}< T <T_{c1}$ the
hexatic order is first established through a KT transition and
one would expect the ordered state to be correspondent to
$\Psi_{i}\approx\Psi_{j}$ for all sites i and j, producing two
degenerate minima for the free energy. So for these range of
temperatures, the above Hamiltonian describes a system with the
symmetry of the 2d-Ising  model and then the transition between
the pure hexatic ($\Psi\neq 0, \Phi=0$)and locked phase (hexatic
plus long range herringbone ) ($\Psi\neq 0, \Phi\neq 0$) should
be Ising-like  at $T_{c2}$ with critical properties of 2d-Ising
model. Thus for $J_{2}<<J_{1}<J_{3}$ the model exhibits an KT
transition at $T_{c1}\sim {\pi J_{1}\over 2}$ and an Ising-like
transition upon decreasing the temperature down to $T_{c2}\sim
2.7J_{2}$ ($k_{B}=1$). For $J_{2} \approx {\pi J_{1}\over
5.4}\approx 0.58J_{1}$, the two transition temperatures turn out
to be equal and so a single transition occurs between disordered
and locked phase in which two orderings would be established
simultaneously.

For $ 0.58J_{1}< J_{2} <J_{3} $ the herringbone order would
establish first and cause the correspondent field $\Phi$ to take
nearly the same value for all sites. Because of this, the
coupling term $J_{3}$ acts like a field on $\Psi$ and so the
hexatic order parameter takes a nonzero value. So for this range
of coupling constants the long range herringbone ordering will
induce hexatic order at the same time.

To obtain a qualitative picture of transitions and also the
approximate location of the critical points, we first set a low
resolution simulations. The Simulations were carried out using
standard Metropolis spin-flipping algorithm with lattice size
$L=20 \times 20$. During each simulation step, The angles
$\Psi_{i}$ and $\Phi_{i}$ were treated as unconstrained,
continuous variables. The random-angles rotations
($\Delta\Psi_{i}$ and $\Delta\Phi_{i}$) were adjusted  in such a
way that roughly $50\%$ of the attempted angle rotations were
accepted. To ensure thermal equilibrium, 100 000 Monte Carlo
steps (MCS) per spin were used for each temperature and 200 000
MCS were used for data collection. The basic thermodynamic
quantities of interests are the specific heat $c=(\langle E^2
\rangle-\langle E \rangle^{2})/(T^{2}L^{2})$, the herringbone
order parameter
$M=L^{-2}[(\sum_{i}\cos(\Phi_{i}))^2+(\sum_{i}\sin(\Phi_{i}))^2]$
and the susceptibility $\chi=(\langle M^2 \rangle-\langle M
\rangle^{2})/(TL^{2})$.

We have obtained the specific-heat,susceptibility and order
parameter data as a function of temperature, shown in Fig. (1-3)
for $J_{1}=1.0$ and $J_{2}=0.5,0.6,0.8,1.0$ and $J_{3}=2.0$.
 From the preceding
discussion, it is clear that the small broad peaks near $T=1.1$ in
figures (1) and (2)  signal the XY transition due to the $J_{1}$
term, while the sharp peak located at $T\sim 0.9$ is expected to
signal a transition into the state of 2d-Ising  symmetry. For
$J_2>0.6$, as already mentioned, only one sharp peak is observed
in the specific-heat and susceptibility data  which verifies our
argument that for these values of $J_{2}$, the transition from
disordered to long range herringbone ordered phase, simultaneously
induces hexatic ordering.

\subsection{Wolff's embedding trick}
One important problem in Monte Carlo simulation, especially for
large systems, is critical slowing down which is a major source
of errors in measurements. To overcome the critical slowing down
we used Wolff's embedding technique{\cite{wolf}}. This method is
based on the cluster algorithm which originally proposed by
Swendsen and wang for the potts model\cite{sw}. In Swendsen and
Wang cluster algorithm a configuration of activated bonds is
constructed from the spin configuration and after the clusters of
spins are formed from configuration of bonds, the spin
configuration is updated by assigning  a randomly  new spin value
to each cluster and then the same value is given to all spins in
the same cluster.

Wolff suggested a  single-cluster algorithm in the way that only a
single cluster grows from a randomly chosen site and then all the
spins in the cluster are flipped. This single-cluster algorithm is
very successful when applied to the Ising model\cite{tamayo}.

 Wolff further developed his technique For spin systems with continuous symmetry by introducing
 the Ising variable into $O(n)$ ferromagnetic models. Choosing a direction in the
spin space at random each spin is projected onto that direction
with two components, one perpendicular and the other either
parallel or anti-parallel to the randomly chosen direction. An
Ising model is then constructed by assigning $+1$ to to spins of
parallel components and $-1$ to spins of anti-parallel
components. The couplings between the nearest-neighbor Ising
spins are determined by the products of these parallel and
anti-parallel components and are therefore random in magnitude
but are ferromagnetic. Such a random-bond Ising model can
efficiently simulated with the single-cluster algorithm and the
original $O(n)$ model can be correspondingly updated by changing
the sign of parallel or anti-parallel components of spins in the
same cluster\cite{kandel,chen,peczak}.

For cluster updating of the coupled $XY$ model, we performed the
following steps:

i)Choose a random oriented direction in 2-dimensional  system with
angle $\theta$ respect to $x$-axis and find the relative rotation
of one of the two fields, i.e hexatic field ($\Psi$), respect to
the randomly chosen direction ($\Psi'_{i}=\Psi_{i}-\theta$);

ii)For each axis direction, generate independent random-bond Ising
models for $\Psi$ variables by assigning $+1$ to each lattice
point if $\cos(\Psi'_{i})>0$ and $-1$ if $\cos(\Psi'_{i}) <0$ ;

iii) For each resultant random-bond Ising model correspondent to
hexatic field, choose a lattice site randomly and build a single
cluster with a bond-activating probability

\begin{equation}
P_{ij}=1-\exp(min(0,-2K_{1}\cos(\Psi')_{i}\cos(\Psi')_{j}),
\end{equation}

where $K_{1}=J_{1}/T$  and the Boltzamann constant being 1;

iv) The spins in each cluster feel the effect of $\Phi$ fields
through coupling term $J_{3}$. Once the $\Psi$ cluster is formed
update the its configuration by flipping all correspondent
embedded Ising variables in each  cluster using Metropolis
algorithm . For this purpose consider $\Delta E$ as energy
difference of spin flipped and initial configurations for a given
cluster in such a way that if $\Delta E < 0$ all spins in the
cluster will be flipped ($\Psi'_{i}\rightarrow\pi-\Psi'_{i}$) and
if $\Delta E>0$ they will be flipped by probability
$p=\exp(-K_{3}\Delta E$ in which $K_{3}=J_{3}/T$. Note that in
this procedure all $\Phi$ variables remain unchanged.

v) Repeat (ii) to (iv) several times before going to next step;

vi)Now fix $\Psi$ variables and repeat steps (ii) to (v) for
$\phi$ variables, Noting that in step (iii) $K_{1}$ should change
to $K_{2}=J_{2}/T$.

vii) Turn to step (i) and choose a new random direction.

This multiple-updating scheme satisfies  detailed balance  and
ergodicity and critical slowing down  is reduced dramatically. To
improve the quantity of data we combined the above algorithm with
the single flip Metropolis method.

All simulations were carried out at five temperatures close to the
effective transition points of the square lattices  with linear
sizes $L=20,24,28,32,36,40,50$ and $60$, by characterizing the
corresponding peak position of the specific-heat and
finite-lattice susceptibility. In each simulation $1\times 10^5$
cluster updating runs were carried out for equilibration. For
data collection, $4\times 10^5$ measurements were made after
enough single cluster-updating followed by single flip Metropolis
runs  are skipped (at least 10) to reduce the correlation between
the measurements. Values of total energy and magnetization from
each measurement were stored as a data list for histogram
analysis.

\subsection{Histogram Method}
To determine The location of the transition temperatures and other
thermodynamic quantities such as specific heat near the
transition points we need to use high resolution methods. For
this purpose we used  multiple-histogram reweighting method
proposed by Ferrenberg and Swendsen {\cite{fs}}, which makes it
possible to obtain accurate data over the transition region from
just a few Monte Carlo simulations.
 The central idea behind the histogram method is to build up
information on the energy probability density function
$P_{\beta}(E)$, where $\beta=1/T$ is inverse temperature (in units
with $k_{B}=1$). A histogram $H_{\beta}(E)$ which is the number of
spin configurations generated between $E$ and $E+\delta E$.
$P_{\beta}(E)$ is  defined as :

\begin{equation}
 P_{\beta}(E_{i})=\frac{H_{\beta}(E_{i})}{Z_{\beta}},
\end{equation}

where

\begin{equation}
 Z_{\beta}=\sum_{i} H_{\beta}(E_{i}).
\end{equation}

On the other hand we now that $P_{\beta}(E_i)$ is proportional to
the Boltzmann weight $\exp(-\beta E_{i})$ as:

\begin{equation}
P_{\beta}(E_{i})=\frac{g(E_{i})\exp(-\beta E_{i})}{Z_{\beta}},
\end{equation}

in which $g(E_{i})$ is the density of states with energy $E_{i}$
and is independent of temperature. By knowing the probability
distributions in a specific temperature, we can derive the
density of states and find the probability distribution of energy
at any temperature $\beta^{'}$ as follows:

\begin{equation}
P_{\beta^{'}}(E_{i})=\frac{P_{\beta}(E_{i})\exp[(\beta-\beta^{'})E_{i}]}
{\sum_{j}P_{\beta}(E_{j})\exp[(\beta-\beta^{'})E_{j}]}.
\end{equation}

In principle, $P_{\beta}(E)$ only provides information on the
energy distribution of nearby temperatures. This is because the
counting statistics in the wings of the distribution
$H_{\beta}(E)$, far from the average energy at temperature $T$,
will be poor.

To improve the estimation for density of states, one can  take
data at more than one temperature and combine the resultant
histograms so as to take the advantages of the regions where each
provide the best estimate for the density of states. This method
has been studied by Ferrenberg and Swendsen who presented an
efficient way for combining the histograms {\cite{fs}}. Their
approach relies on first determining the characteristic
relaxation time $\tau_{j}$ for the $j$th simulation and using
this to produce a weighting factor $g_{j}=1+2\tau_{j}$. The
overall probability distribution at coupling $K=\beta J$ obtained
from $n$ independent simulation, each with $N_{j}$
configurations, is then given by :

\begin{equation}\label{mh}
P_{K}(E)=\frac{[\sum_{j=1}^{n}g_{j}^{-1}H_{j}(E)]e^{-KE}}
{\sum_{j=1}^{n}N_{j}g_{j}^{-1}e^{-K_{j}E-f_{j}}},
\end{equation}

where $H_{j}(E)$ is the histogram for the $j$th simulation and
the factors $f_{j}$ are chosen self-consistently using
Eq.(\ref{mh}) and

\begin{equation}
e^{f_{j}}=\sum_{E}P_{K_{j}}(E).
\end{equation}

Thermodynamic properties are determined, as before, using this
probability distribution, but now the results would be valid over
a much wider range of temperatures than for any single histogram.
In addition, this method gives an expression for the statistical
error of $P_{K}(E)$ as:

\begin{equation}
\delta P_{K}(E)=[\sum_{j=1}^{n}g_{j}^{-1}H_{j}(E)]^{-1/2}P_{K}(E),
\end{equation}

from which it is clear that the statistical error will be reduced
when more MC simulations are added to the analysis.

To deal with thermodynamic quantities other than the energy, one
can choose to to work with energy probability distribution and
microcanonical averages of the quantities of interest. This leads
to optimized use of computer memory. The microcanonical average of
a given quantity $M$, which is a function of energy,  can be
calculated directly  as :

\begin{equation}
M(E)=\frac{\sum_{t}M_{t}\delta_{E_{t},E}}{\sum_{t}\delta_{E_{t},E}},
\end{equation}

from which the canonical average of $M$ can be obtained as a
function of $T$:

\begin{equation}
\langle M \rangle=\frac{\sum_{E}M(E)P(E,T)}{\sum_{E}P(E,T)}.
\end{equation}

The temperature dependence of thermodynamics quantities were
determined by optimized multiple-histogram method. For all system
sizes, histograms obtained from simulations overlap sufficiently
on both sides of the critical point so that the statistical
uncertainty in the wing of the histograms, near the critical
point may be suppressed by using the optimized multiple-histogram
method. Therefore the locations and magnitudes of the extrema of
the thermodynamic quantities can be determined accurately to
extract the critical temperature and static critical exponents
from the finite-size scaling behaviour.

\subsection{Determination of $T_{c}$ and static critical exponents}

In order to determinate the critical temperature in the infinite
lattice sizes as well as the critical exponents, we use the
finite-size scaling theory \cite{barber}. According to the
finite-size scaling theory, the free energy density can be
divided into a singular part $f_{s}$ and a background $f_{ns}$
which is non-singular as :

\begin{equation}
f(t,h;L)=f_{s}(t,h;L)+f_{ns}(t,L),
\end{equation}

where $t=(T-T_{c})/T_{c}$ is the reduced temperature for a
sufficiently large system at a temperature $T$ close enough to
the infinite lattice critical point $T_{c}$ and $h$ is the
external ordering field . Using the periodic boundary conditions
makes The non-singular part size independent, leaving only the
singular part of free energy for studying the critical properties
of the system. The singular part is described phenomenologically
by a universal scaling form

\begin{equation}
f(t,H;L)=L^{-d}Y(tL^{y_{t}},hL^{y_{h}})+\cdot\cdot\cdot,
\end{equation}

where $d$ is the spatial dimension of the system and $y_{t}$ and
$y_{h}$ are related to static critical exponents as $y_{t}=1/\nu$
and $y_{h}={\beta\delta\over\nu}$. Scaling form for various
thermodynamic quantities can be obtained  from proper derivations
of the free energy density. Some of them such as magnetization
density, susceptibility and specific heat in zero field are:

\begin{eqnarray}
\label{mag}m&\approx& L^{\beta/\nu}{\mathcal{M}}(tL^{1/\nu})\\
\label{kappa}\chi&\approx& L^{\gamma/\nu}{\mathcal{K}}(tL^{1/\nu})\\
\label{sh}c&\approx&
c_{\infty}(t)+L^{\alpha/\nu}{\mathcal{C}}(tL^{1/\nu})
\end{eqnarray},

where $\alpha$ ,$\beta$,$\gamma$ and $\delta$ are static critical
exponents. The Eqs(\ref{mag}-\ref{sh}) are used to estimate the
critical exponents. But before dealing with the critical
exponents  we should first determine the critical temperature
accurately.

The logarithmic derivatives are total magnetization ($mL^{d}$) are
important thermodynamic quantities for studying critical
phenomena and very useful to high accurate estimation of the
critical temperature $T_{c}$ and the critical exponent $\nu$
\cite{chen}. For example defining the following quantities:

\begin{eqnarray}
\label{v1}V_{1}&\equiv& 4[M^{3}]-3[M^{4}],\\
V_{2}&\equiv& 2[M^{2}]-[M^{4}],\\
V_{3}&\equiv& 3[M^{2}]-2[M^{3}],\\
V_{4}&\equiv& (4[M]-[M^{4}])/3,\\
V_{5}&\equiv& (3[M]-[M^{3}]/2,\\
\label{v6}V_{6}&\equiv& 2[M]-[M^{2}],
\end{eqnarray}

where $M=mL^{d}$ is the total magnetization of the system and

\begin{equation}
[M^{n}]\equiv \ln\frac{\partial\langle M^{n} \rangle}{\partial T}.
\end{equation}

From Eq. (\ref{mag}) it is easy to show that

\begin{equation}
\label{vj}V_{j}\approx (1/\nu)\ln L+{\mathcal{V}}_{j}(tL^{1/\nu})
\end{equation}

For $j=1,2,\cdot\cdot\cdot,6$. At the critical temperature
($t=0$) the ${\mathcal{V}}_{j}$ should be constants independent of
the system size $L$. As we will see it gives us a very accurate
tool to estimate both the critical temperature and correlation
length exponent $\nu$ independently with high precision.

\subsection{Order of the transition}

To determine the order of transitions, we used Binder's forth
energy cumulant defined as:

\begin{equation}
 U_{L}=1-\frac{<E^4>}{3<E^2>}.
\end{equation}

It has been shown that this quantity reaches a minimum at the
effective transition temperature $T_{c}(L)$ whose  size
dependence is given by:

\begin{equation}\label{bind}
U_{min}(L)=U^{*}+BL^{-d}+O(L^{-2d}),
\end{equation}

where

\begin{equation}
U^{*}=\frac{2}{3}-\left(e_{1}/e_{2}-e_{2}/e_{1}\right)^{2}/12.
\end{equation}

The quantities $e_{1}$ and $e_{2}$ are the values of energy per
site at the transition point a first order phase transition and
$d$ is the spatial dimension of the system ($d=2$ in our
simulation). Hence, for the continuous transitions for which
there is no latent heat ($e_{1}=e_{2}$), in the limit of infinite
system sizes, $U_{min}(L)$ tends to the value $U^{*}$ equal to
$2/3$. For the first-order transitions, however $e_{1}\neq e_{2}$
and then $U^{*}$ reaches a value less than $2/3$ in the the limit
$L\rightarrow\infty$. This method is actually a test for the
Gaussian nature of the probability density function $P(E)$ at
$T_{c}$. For a continuous transition, $P(E)$ is expected to be
Gaussian at, as well as away from $T_{c}$. For a first-order
transition, $P(E)$ will be double peaked in infinite lattice size
limit, hence deviation from being Gaussian cause the minimum of
$U_{L}$ tends $U^{*}$ to be less than $2/3$ as
$L\rightarrow\infty$. This method is very sensitive, in a sense
that small splitting in $P(E)$ for the infinite system that do
not result in a double peak for small lattices can be detected.

\section{results and discossion}

we are interested in investigating  those region in coupling
constants space for which the two kinds of ordernig establish
together, so we limit ourselves to ${J_{2}\over J_{1}}> 0.6$.
Fixing $J3=1.0$ and $J_{3}=2.0$ we start to get data for
$J_{2}=0.6,0.7,0.8,0.9,1.0$.

First of all we deal with the order of transitions. Employing the
forth Binder energy cumulant method, discussed in previous
section, we found that the order of transition for all values of
$0.6 \leq J_{2}\leq 1.0$ are second order in contrast to
\cite{gk}. For example, we have plotted the size dependence of
minimum values of $U_L$ ($U_min(L)$ vs $L^{-2}$) for
$J_{2}=0.6,0.8$ and $1.0$ in Fig.(4). As can be seen from this
figure, the asymptotic values $U^{*}$ for all the $J_{2}$'s are
equal to $2/3$ in the measurement errors, which indicate the
continuous transition for these range of couplings.

After determining the order of transitions we proceed to estimate
the critical temperatures and the critical exponents, using the
finite-size scaling. The analysis discussed in section II.D
provides a way to simultaneously determination of both $\nu$ and
$T_{c}$. For this purpose, using Eq. (\ref{vj}) one can find the
slope of quantities $V_{1}$ to $V_{6}$ (Eq. \ref{v1}-\ref{v6})
versus $\ln(L)$ for the region near the critical point. Scanning
over the critical region and looking for a quantity-independent
slope gives us both the critical temperature $T_{c}$ and
correlation length exponent $\nu$ with high precision. Figures
(5) and (6) give the  examples of such an effort  for the set of
the couplings ($J1=1.0, J2=0.8,J3=2.0$). From both figures we
estimate that $\nu=0.837(5)$ and $T_{c}=1.157(1)$. The linear fits
to the data in figure (5) has been obtained by the linear least
square method. The similar procedure for couplings set ($J1=1.0,
J2=1.0,J3=2.0$), shown in figures (7)  gives $\nu=1.01(3)$ and
$T_{c}=1.051(1)$.

Once $\nu$ and $T_{c}$ are determined  accurately, we can extract
other static critical exponents related to the order parameter
($\beta$) and susceptibility ($\gamma$). The ratio $\beta/\nu$
can be estimated by using the size dependence of the order
parameter at the critical point given by Eq.(\ref{mag}). Figures
(8) and (9) are  log-log plots of the size dependence of the order
parameter corresponding to field $\Phi$ for $J_{2}=0.6$ and
$J_{2}=0.8$ respectively. From these figures  the ratio
$\beta/\nu$ can be estimated as the slope of the straight lines
fitted to the data according to Eq.(\ref{mag}). We then have
$\beta/\nu=.143(8)$ for $J_{2}=0.6$ and $\beta/\nu=.169(6)$ for
$J_{2}=0.8$ and therefore $\beta=0.144(8)$ for $J2=0.6$ and
$\beta=0.142(5)$ for $J_{2}=0.8$.

Accordingly, from Eq.(\ref{kappa}) it is clear that the peak
values of the finite-lattice susceptibility ($\chi=(\langle M^2
\rangle-\langle M \rangle^{2})/(TL^{2})$) and the magnitude of the
true susceptibility at $T_{c}$ (the same as $\chi$ with $\langle m
\rangle=0$)  are asymptotically proportional to $L^{\gamma/\nu}$
. So the slope of straight line fitted linearly to the log-log
plot of these two quantities versus linear size of the lattices,
can be calculated to estimate the ratio $\gamma/\nu$. Such plots
have been depicted in Figures.(10) and (11) for $J2=0.6$ and
$J2=0.8$, respectively. In Fig.(10), the slope of the bottom
straight line (finite-lattice susceptibility) is $1.726(8)$ from
the linear fit and the slope for the top one is $1.709(7)$, where
the error includes the uncertainty in the slope resulting from
uncertainty in our estimate for $T_{c}$. The ratio $\gamma/\nu$
obtained from the average of two slopes is $1.717$, therefore
knowing the value of $\nu=1.01(3)$ one gets $\gamma=1.73(6)$ for
$J2=0.6$ and $J3=2.0$. Similarly for $J_{2}=0.8$, depicted in
Fig.(11), the slope of the bottom straight line (finite-lattice
susceptibility) is $1.637(5)$ from the linear fit, while for the
top one is $1.656(7)$, so he ratio $\gamma/\nu$ obtained from the
average of the slopes is $1.656(9)$ which results in
$\gamma=1.39(1)$ for $J2=0.8$ and $J3=2.0$.

Above procedure has been applied for other values of $J_{2}$ and
the critical temperatures and critical exponents and for all
values of herringbone couplings obtained in this work, have been
listed in table.(1). In this table, the critical exponents
$\alpha$, $\delta$ and $eta$ have been calculated using scaling
laws:
\begin{eqnarray}
\alpha&=&2-d\nu \\
\gamma&=&\nu(2-\eta)=\beta(\delta-1)\\
\end{eqnarray}

on the other hand the Rushbrook scaling law
($\alpha+2\beta+\gamma=2$) is satisfied for all set of exponents
in the computation errors.

In order to check the values of specific-heat anomaly exponents,
we also estimated them independently, using size dependence of
the specific-heat at measured critical temperatures according to
Eq.(\ref{sh}). For this purpose we used least square method to
find the best value of $\alpha/\nu$ to fit the heat capacity data
at $T_{c}$. The plot of such efforts have been represented  in
figures () and () for $J_{2}=0.7$ and $J_{2}=0.8$ respectively,
which resulted in $\alpha/\nu=45(2)$ for $J_{2}=0.7$ and
$\alpha/\nu=0.39(1)$ for $J_{2}=0.8$. So knowing  the obtained
values of $\nu$, we find $\alpha=0.36(2)$ for $J_{2}=0.7$ and
$\alpha=0.33(1)$ for $J_{2}=0.8$. This results are in very good
agreement with the values obtained from Josefson scaling law.

One can see from table.(1) that the critical exponents for
$J_{2}=0.6$ are close to two dimensional Ising model for which the
exact exponents are
$\alpha=0,\nu=1.0,\beta=0.125,\gamma=1.75,\delta=15$ and
$\eta=0.25$. So $J_{2}=0.6$ is the onset of 2d-ising behaviour.
By increasing the herringbone coupling constant, the thermal
exponents $\alpha$ and $\nu$ differ dramatically from that of
2d-ising value. Heat capacity anomaly exponent $\alpha$ gets its
maximum value ($\alpha=0.39$) for $J_{2}=0.7$ and then decreases
to $0.22$ for $J_{2}=1.0$. The critical exponents for $J_{2}=0.9$
and $J_{2}=1.0$ are equal up to the simulation errors, hence these
two belong to the same universality class.

The specific-heat exponent corresponding to $J_{2}=0.8$
($\alpha=0.32(2)$) is the closest one to experimentally observed
values ($\alpha=0.31\pm 0.03$). To investigate to sensibility of
the critical exponents to hexatic-herringbone couplings we also
measured the critical exponents for the coupling set
$J_{1}=1.0,J_{2}=0.8$ and $J_{3}=1.0$. The static critical
exponents obtained for this set of couplings are as
$\alpha=0.31(3),\nu=0.845(15),\beta=0.13(2),\gamma=1.43(3),\delta=12.2(1.9)$
and $\eta=0.30(1)$. As we see again the heat capacity exponent is
in excellent agreement with experimental value.

\section{Conclusion }

In summary, using the optimized Monte Carlo simulation based on
multi-histogram and wolf's embedding  methods, we investigated the
critical properties  associated with a Hamiltonian containing two
coupled XY order parameters (indicating a hexatic field with
sixfold symmetry and a long-ranged herringbone field with twofold
symmetry) in two dimensions. Unlike  Bruinsma and Aeppli model,
which possess the three state potts symmetry, the model proposed
in this paper has 2d-Ising symmetry. However, static critical
exponents derived by finite-size analysis for some range a
coupling in coupling constants space, indicates clear deviations
from two dimensional Ising behaviour in transition between
isotropic to hexatic+herringbone phases. Our results show a
non-universal characteristics  along the transition line on which
the two kinds of orderings would establish simultaneously. For the
value of  herringbone coupling ($J_{2}/J_{1}=0.6$), which is the
onset of only one transition from disorder to locked phase, the
critical behaviour is Ising-like, while by increasing the
herringbone coupling the critical exponents begin to deviate from
those of Ising values and finally reach to a new universality
class correspond to $J_{2}/J_{1}=0.9$ and $J_{2}/J_{1}=1.0$ for
which the exponents remain unchanged. Surprisingly for some
values of herringbone couplings in between these to limit, i.e
$J_{2}/J_{1}\sim 0.8$ and $J_{3}/J{1}=2.0,1/0$, the heat capacity
exponents show very good agreement with experimentally observed
values. Whether or not that these transitions are indicating a
new universality or are just a crossover behaviour is an open
problem and requires more theoretical investigations based on
renormalization group theory.

Our results suggest that coupling of hexatic ordering to
long-range herringbone packing but short-range translational
ordering could give a plausible description for large
specific-heat anomaly exponents of SmA-HexB transition in some
liquid crystal compounds for which the herringbone ordering is
accompany with the hexatic ordering. The confirmation of this
idea requires  the similar simulation in three dimensions and is
the subject of our current research.

Experimentally, the accurate measurements of the range of
herringbone ordering in HexB phase of $nm$OBC compounds in bulk
and thin film samples and also measuring other static critical
exponents rather than heat capacity exponent in thin film samples
are also needed to check the validity of this model.

We finally hope that our work will motivate further theoretical,
numerical and experimental investigations of this very
interesting problem.

{\bf Acknowledgment} We would like to thank prof. Hadi Akbarzadeh
for letting us to use his  computational facilities.

%\newpage

%%%%%%%%%%%%%%%%%%%%%%%%%%%%%%%%%%%%%%%%%%%%%%%%%%%%%%%%%%%%%%%%%%
\begin{table}[t]
\begin{tabular}{|c|c|c|c|c|c|c|c|} %\hline
$J_{2}/J_{1}$ &  $T_{c}$ & $\nu$ & $\beta$ & $\gamma$ & $\alpha$ & $\delta$ & $\eta$ \\
\hline
 0.6    &  1.051(1)    &  1.01(3) & 0.144(14)  &  1.73(6) & -0.02(6) & 13.0(1.6) & 0.282(8)  \\
 0.7    &  1.164(1)    &  0.806(9) & 0.143(9)  &  1.31(3) & 0.39(2) & 10.1(9) & 0.370(30)  \\
 0.8    &  1.257(1)    &  0.837(8) & 0.142(5)  &  1.39(1) & 0.32(2) & 10.8(4) & 0.344(9)  \\
 0.9    &  1.342(1)    &  0.886(3) & 0.156(5)  &  1.46(1) & 0.23(1) & 10.3(4) & 0.35(1)  \\
 1.0    &  1.423(1)    &  0.888(3) & 0.150(4)  &  1.48(2) & 0.22(2) & 10.8(4) & 0.33(1)
 %\hline
\end{tabular}
%{\box1}
\narrowtext\caption{The critical temperatures and static critical
exponents for ${J_{2}\over J_{1}}=0.6,0.7,0.8,0.9,1.0$ and
${J_{3}\over J_{1}}=2.0$, derived from finite-size scaling.(see
the text) }
\end{table}
%%%%%%%%%%%%%%%%%%%%%%%%%%%%%%%%%%%%%%%%%%%%%%%%%%%%%%%%%%%%%%%%%%%%%%%%%%%

\newpage
%%%%%%%%%%%%%%%%%%%%%%%%%%%%%%%%%%%%%%%%%%%%%%%%%%%%%%%%%%%%%%%%%%%%%%%%%
\begin{figure}[c]
\epsfxsize=8.5truecm \epsfbox{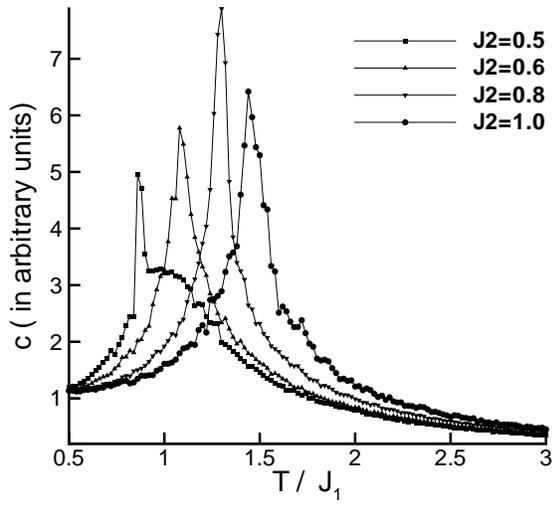} \narrowtext \caption{
Temperature dependence of specific-heat for $J_{1}=1.0$,
$J_{3}=2.0$ and $J_{2}=0.5,0.6,0.8,1.0$ }
\end{figure}
%%%%%%%%%%%%%%%%%%%%%%%%%%%%%%%%%%%%%%%%%%%%%%%%%%%%%%%%%%%%%%%%%%%%%%%%%%%

\newpage
%%%%%%%%%%%%%%%%%%%%%%%%%%%%%%%%%%%%%%%%%%%%%%%%%%%%%%%%%%%%%%%%%%%%%%%%%
\begin{figure}[c]
\epsfxsize=8.5truecm \epsfbox{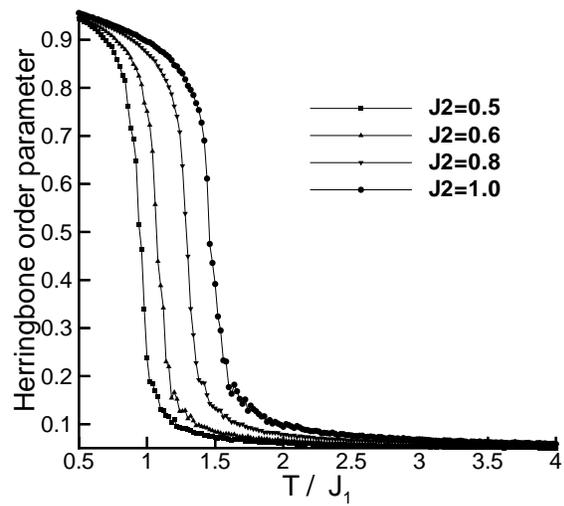} \narrowtext \caption{
Temperature dependence of herringbobe order parameter for
$J_{1}=1.0$, $J_{3}=2.0$ and $J_{2}=0.5,0.6,0.8,1.0$ }
\end{figure}
%%%%%%%%%%%%%%%%%%%%%%%%%%%%%%%%%%%%%%%%%%%%%%%%%%%%%%%%%%%%%%%%%%%%%%%%%%%

\newpage
%%%%%%%%%%%%%%%%%%%%%%%%%%%%%%%%%%%%%%%%%%%%%%%%%%%%%%%%%%%%%%%%%%%%%%%%%
\begin{figure}[c]
\epsfxsize=8.5truecm \epsfbox{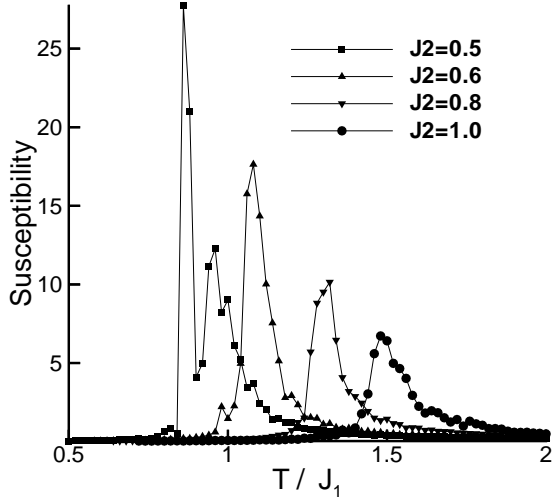} \narrowtext \caption{
Temperature dependence of susceptibility for $J_{1}=1.0$,
$J_{3}=2.0$ and $J_{2}=0.5,0.6,0.8,1.0$ }
\end{figure}
%%%%%%%%%%%%%%%%%%%%%%%%%%%%%%%%%%%%%%%%%%%%%%%%%%%%%%%%%%%%%%%%%%%%%%%%%%%

%%%%%%%%%%%%%%%%%%%%%%%%%%%%%%%%%%%%%%%%%%%%%%%%%%%%%%%%%%%%%%%%%%%%%%%%%
\begin{figure}[t]
\epsfxsize=8.5truecm \epsfbox{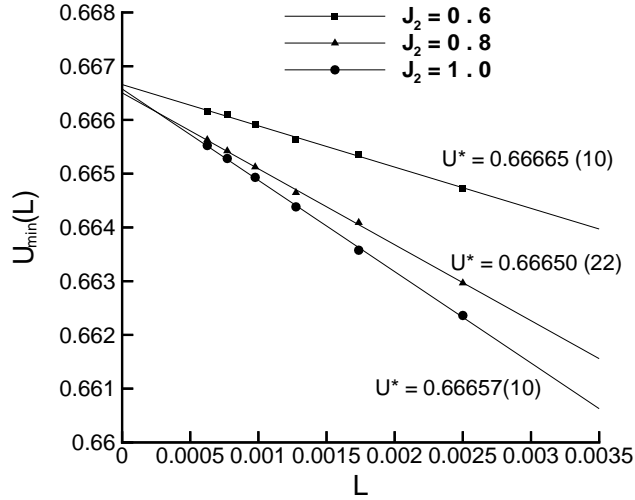} \caption{Size dependence
of binder fourth energy cumulant minima,
 calculated by optimized re-weighting for $J_{1}=1.0$,
$J_{3}=2.0$ and $J_{3}=0.6,0,8,1.0$ . Solid lines represent fits
to (\ref{bind}). Error bars are less than the size of the points.}
\end{figure}
%%%%%%%%%%%%%%%%%%%%%%%%%%%%%%%%%%%%%%%%%%%%%%%%%%%%%%%%%%%%%%%%%%%%%%%%%%%

\newpage
%%%%%%%%%%%%%%%%%%%%%%%%%%%%%%%%%%%%%%%%%%%%%%%%%%%%%%%%%%%%%%%%%%%%%%%%%
\begin{figure}[c]
\epsfxsize=8.5truecm \epsfbox{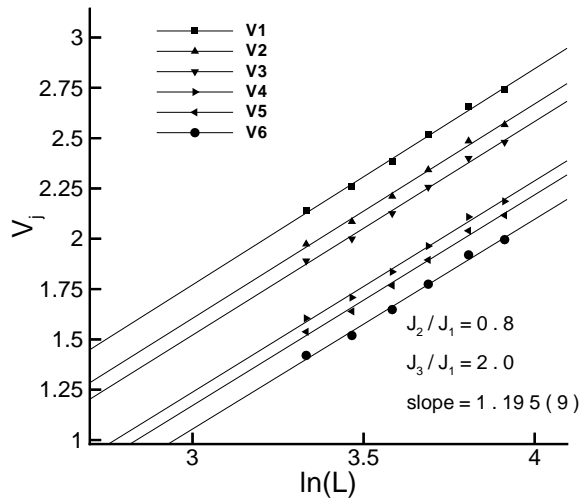} \narrowtext \caption{
dependence of quantity $V_{j}$ (see the text) versus logarithm of
$L$  for $J_{1}=1.0$, $J_{3}=2.0$ and $J_{2}=0.8$ at $T=1.258$.
The solid lines represent linear fits to Eq.(\ref{vj}). All
straight lines have the same slope $1.195$. }
\end{figure}
%%%%%%%%%%%%%%%%%%%%%%%%%%%%%%%%%%%%%%%%%%%%%%%%%%%%%%%%%%%%%%%%%%%%%%%%%%%

\newpage
%%%%%%%%%%%%%%%%%%%%%%%%%%%%%%%%%%%%%%%%%%%%%%%%%%%%%%%%%%%%%%%%%%%%%%%%%
\begin{figure}[c]
\epsfxsize=8.5truecm \epsfbox{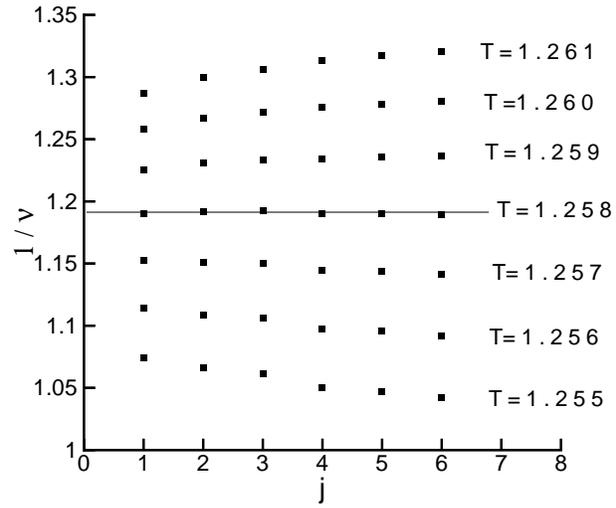} \narrowtext \caption{
Scanning results for the dependence of quantity $V_{j}$  versus
$j$ for $J_{1}=1.0$, $J_{3}=2.0$ and $J_{2}=0.8$. The horizontal
line is drawn at $1/\nu=1.195$. }
\end{figure}
%%%%%%%%%%%%%%%%%%%%%%%%%%%%%%%%%%%%%%%%%%%%%%%%%%%%%%%%%%%%%%%%%%%%%%%%%%%

\newpage
%%%%%%%%%%%%%%%%%%%%%%%%%%%%%%%%%%%%%%%%%%%%%%%%%%%%%%%%%%%%%%%%%%%%%%%%%
\begin{figure}[c]
\epsfxsize=8.5truecm\epsfbox{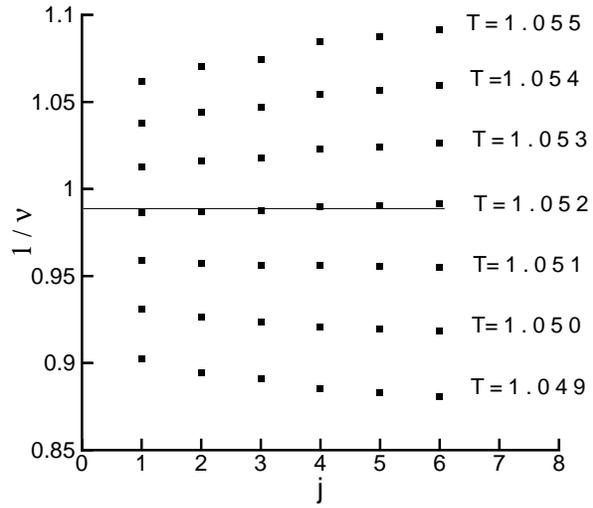}\narrowtext\caption{Scanning
results  of quantity $V_{j}$  for $J_{1}=1.0$, $J_{3}=2.0$ and
$J_{2}=0.6$. The horizontal line is drawn at $1/\nu=0.99$.}
\end{figure}
%%%%%%%%%%%%%%%%%%%%%%%%%%%%%%%%%%%%%%%%%%%%%%%%%%%%%%%%%%%%%%%%%%%%%%%%%%%

\newpage
%%%%%%%%%%%%%%%%%%%%%%%%%%%%%%%%%%%%%%%%%%%%%%%%%%%%%%%%%%%%%%%%%%%%%%%%%
\begin{figure}[c]
\epsfxsize=8.5truecm \epsfbox{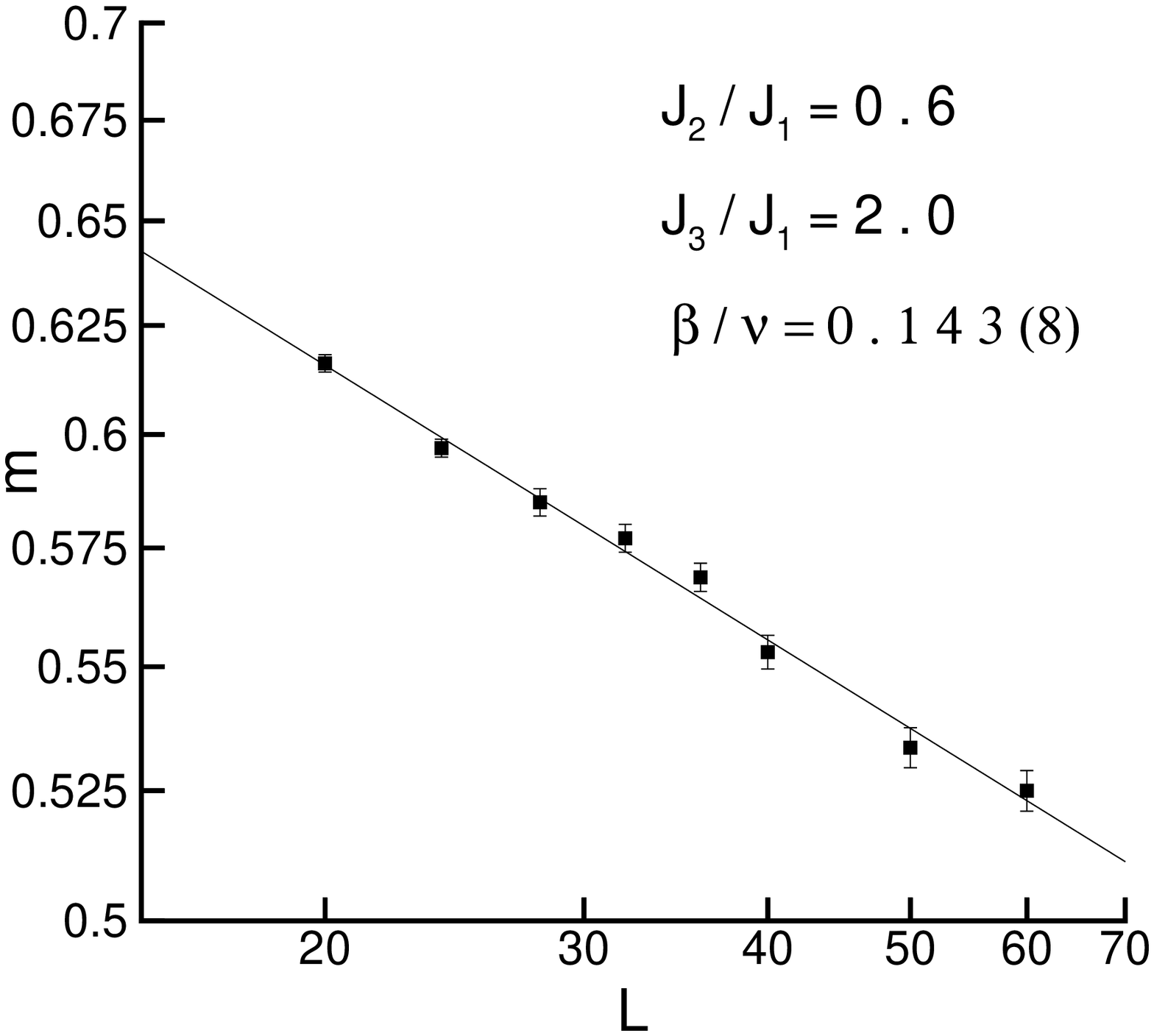} \narrowtext \caption{
Log-log plot of  order parameter versus the linear size of the
lattice $L$ at $T_{c}=1.052$  for $J_{1}=1.0$, $J_{3}=2.0$ and
$J_{2}=0.6$. The solid  line is obtained by  fitting the data to
Eq(\ref{mag}).}
\end{figure}
%%%%%%%%%%%%%%%%%%%%%%%%%%%%%%%%%%%%%%%%%%%%%%%%%%%%%%%%%%%%%%%%%%%%%%%%%%%

\newpage
%%%%%%%%%%%%%%%%%%%%%%%%%%%%%%%%%%%%%%%%%%%%%%%%%%%%%%%%%%%%%%%%%%%%%%%%%
\begin{figure}[c]
\epsfxsize=8.5truecm \epsfbox{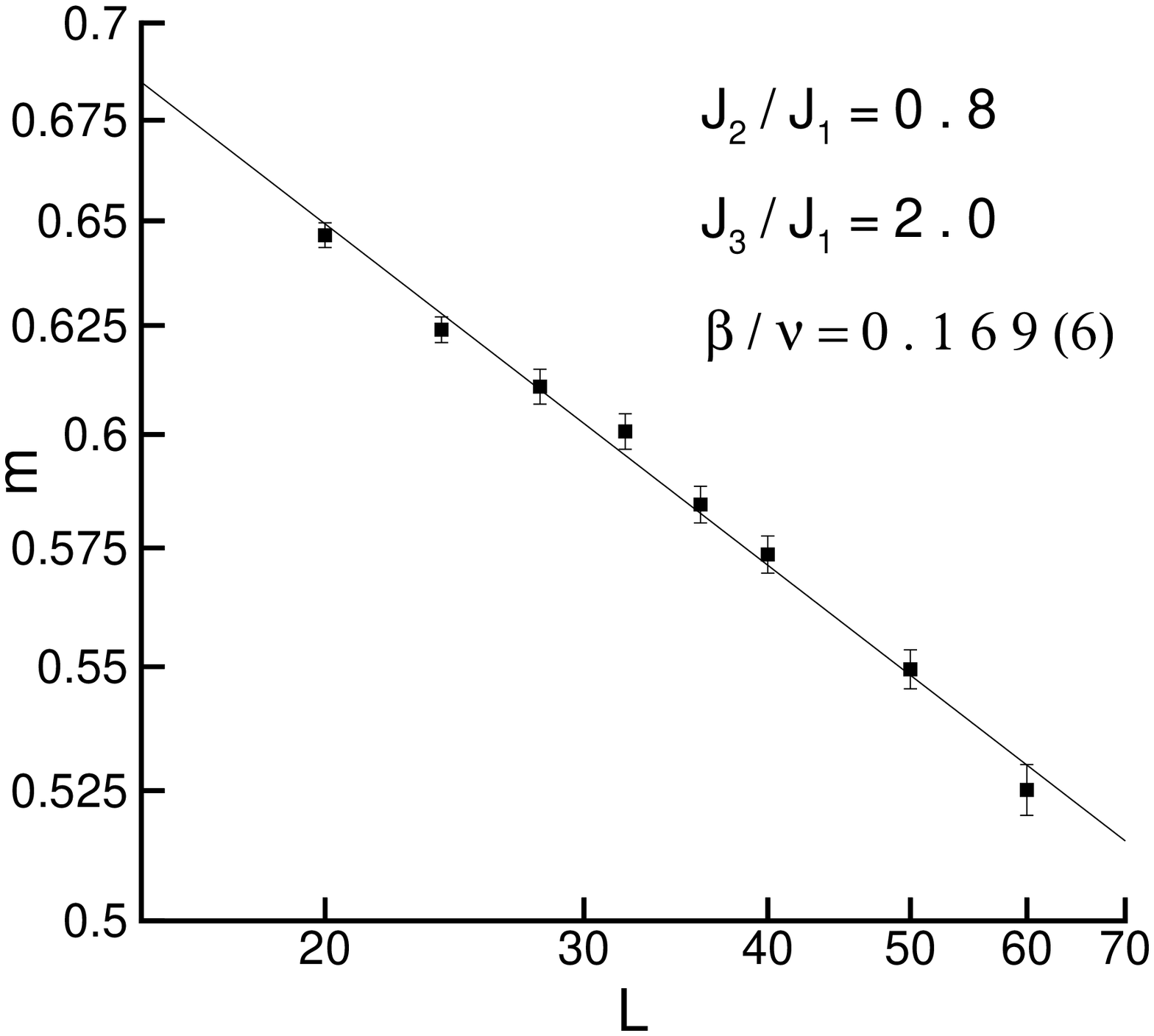} \narrowtext \caption{
Log-log plot of  order parameter versus the linear size of the
lattice $L$ at $T_{c}=1.257$ for  $J_{1}=1.0$, $J_{3}=2.0$ and
$J_{2}=0.8$. The solid  line is obtained by  fitting the data to
Eq(\ref{mag}).}
\end{figure}
%%%%%%%%%%%%%%%%%%%%%%%%%%%%%%%%%%%%%%%%%%%%%%%%%%%%%%%%%%%%%%%%%%%%%%%%%%%

\newpage
%%%%%%%%%%%%%%%%%%%%%%%%%%%%%%%%%%%%%%%%%%%%%%%%%%%%%%%%%%%%%%%%%%%%%%%%%
\begin{figure}[c]
\epsfxsize=8.5truecm \epsfbox{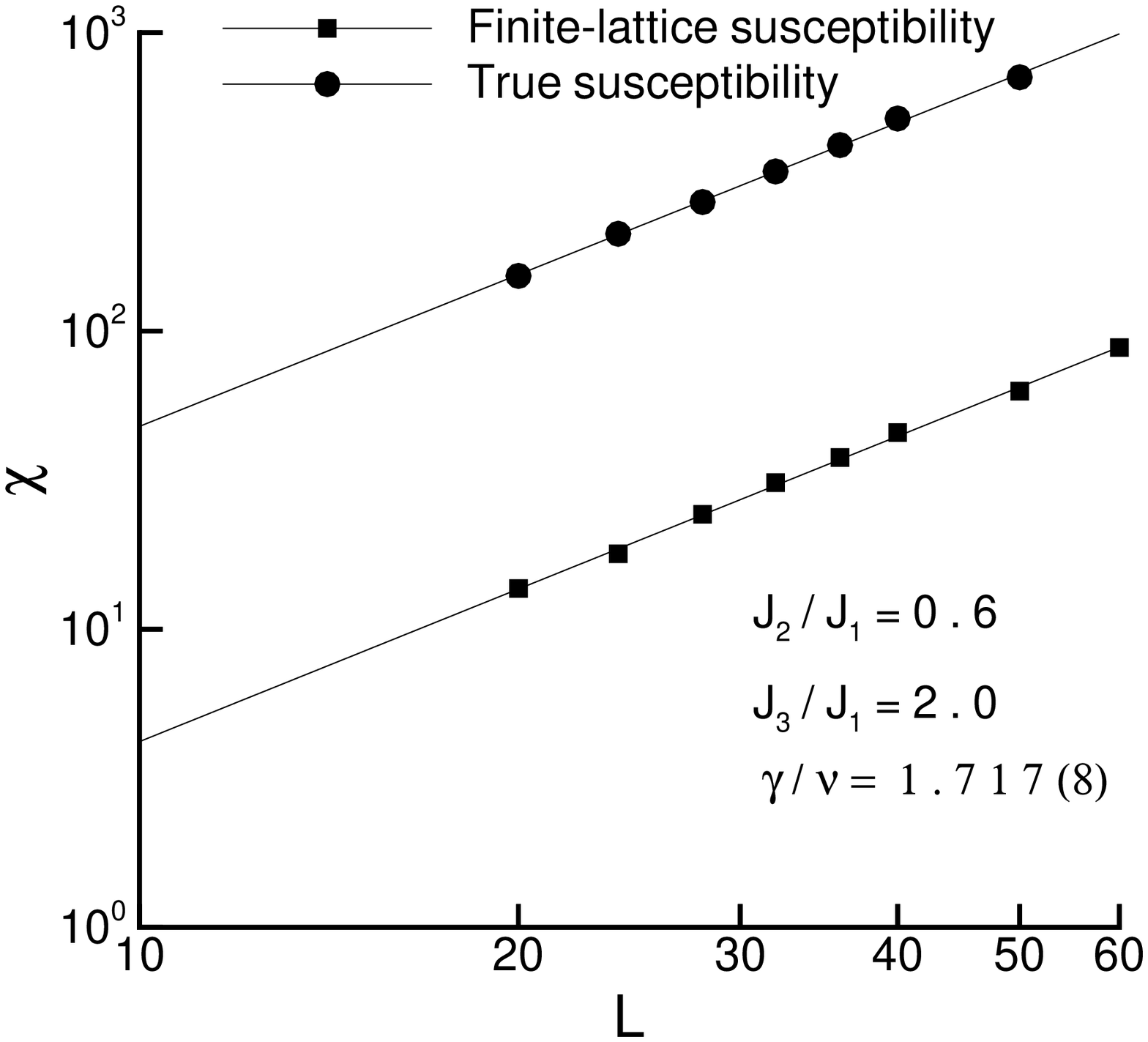} \narrowtext \caption{
Log-log plot of  finite-lattice susceptibility and true
susceptibility versus the linear size of the lattice $L$ at
$T_{c}=1.052$ for $J_{1}=1.0$, $J_{3}=2.0$ and $J_{2}=0.6$. The
solid  line is obtained by fitting the data to Eq(\ref{kappa}).}
\end{figure}
%%%%%%%%%%%%%%%%%%%%%%%%%%%%%%%%%%%%%%%%%%%%%%%%%%%%%%%%%%%%%%%%%%%%%%%%%%%

\newpage
%%%%%%%%%%%%%%%%%%%%%%%%%%%%%%%%%%%%%%%%%%%%%%%%%%%%%%%%%%%%%%%%%%%%%%%%%
\begin{figure}[c]
\epsfxsize=8.5truecm \epsfbox{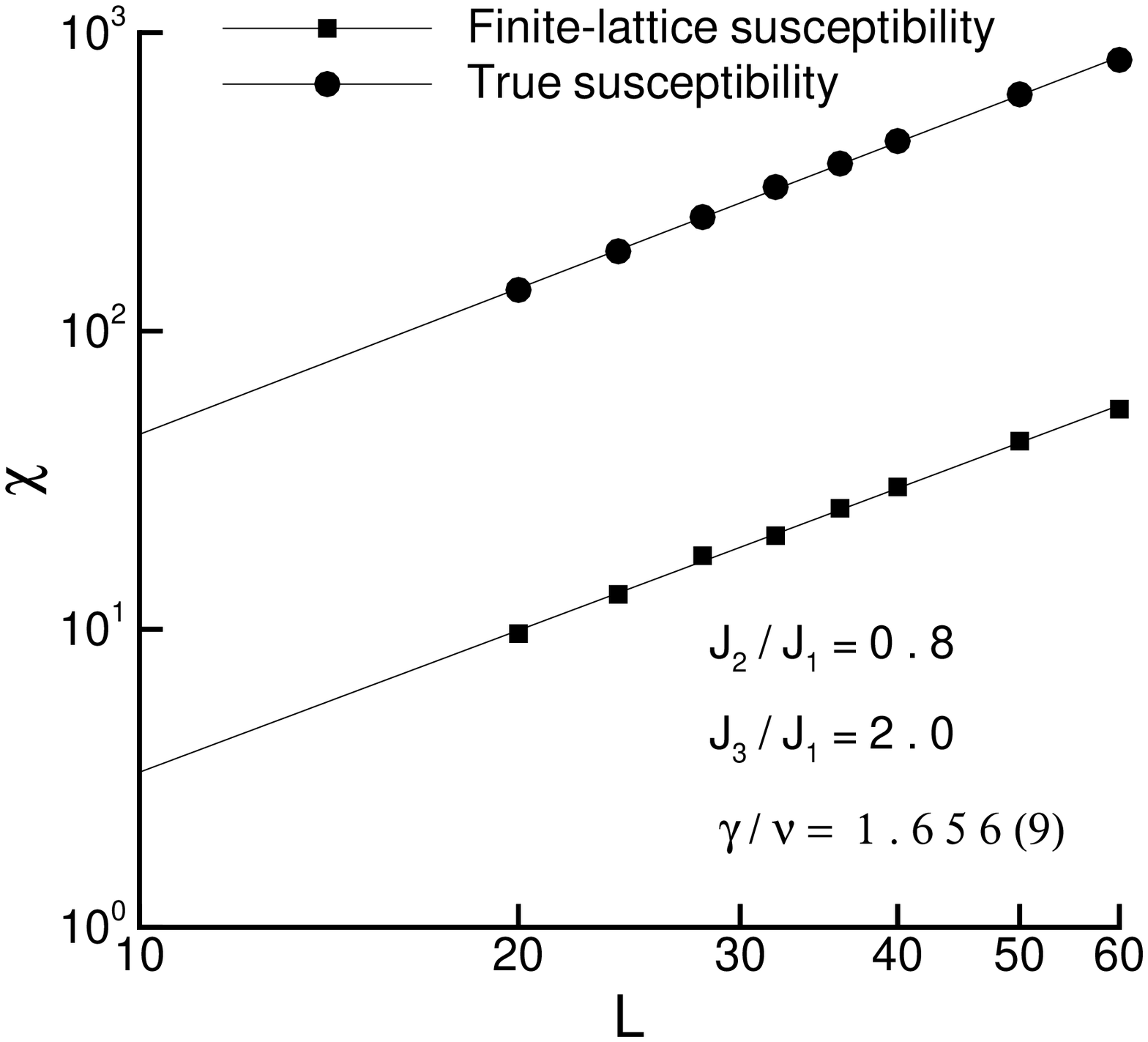} \narrowtext \caption{
Log-log plot of  finite-lattice susceptibility and true
susceptibility versus the linear size of the lattice $L$ at
$T_{c}=1.257$ for $J_{1}=1.0$, $J_{3}=2.0$ and $J_{2}=0.8$. The
solid  line is obtained by fitting the data to Eq(\ref{kappa}).}
\end{figure}
%%%%%%%%%%%%%%%%%%%%%%%%%%%%%%%%%%%%%%%%%%%%%%%%%%%%%%%%%%%%%%%%%%%%%%%%%%%

\newpage
%%%%%%%%%%%%%%%%%%%%%%%%%%%%%%%%%%%%%%%%%%%%%%%%%%%%%%%%%%%%%%%%%%%%%%%%%
\begin{figure}[c]
\epsfxsize=8.5truecm \epsfbox{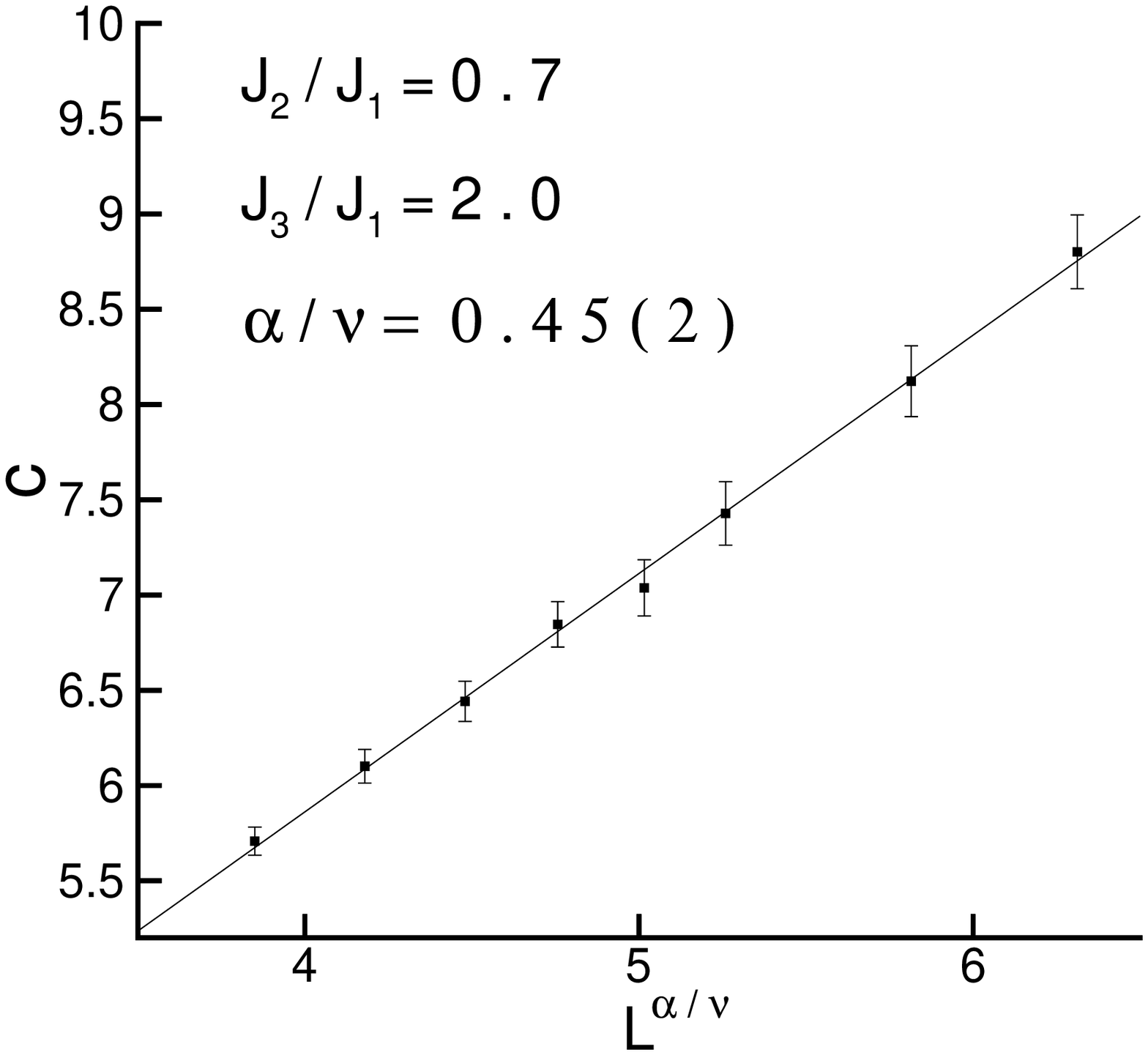} \narrowtext \caption{
Size dependence of specific-heat  at $T_{c}=1.164$ for
$J_{1}=1.0$, $J_{3}=2.0$ and $J_{2}=0.7$. The solid  line is
obtained by fitting the data to Eq(\ref{sh}).}
\end{figure}
%%%%%%%%%%%%%%%%%%%%%%%%%%%%%%%%%%%%%%%%%%%%%%%%%%%%%%%%%%%%%%%%%%%%%%%%%%%

\newpage
%%%%%%%%%%%%%%%%%%%%%%%%%%%%%%%%%%%%%%%%%%%%%%%%%%%%%%%%%%%%%%%%%%%%%%%%%
\begin{figure}[c]
\epsfxsize=8.5truecm \epsfbox{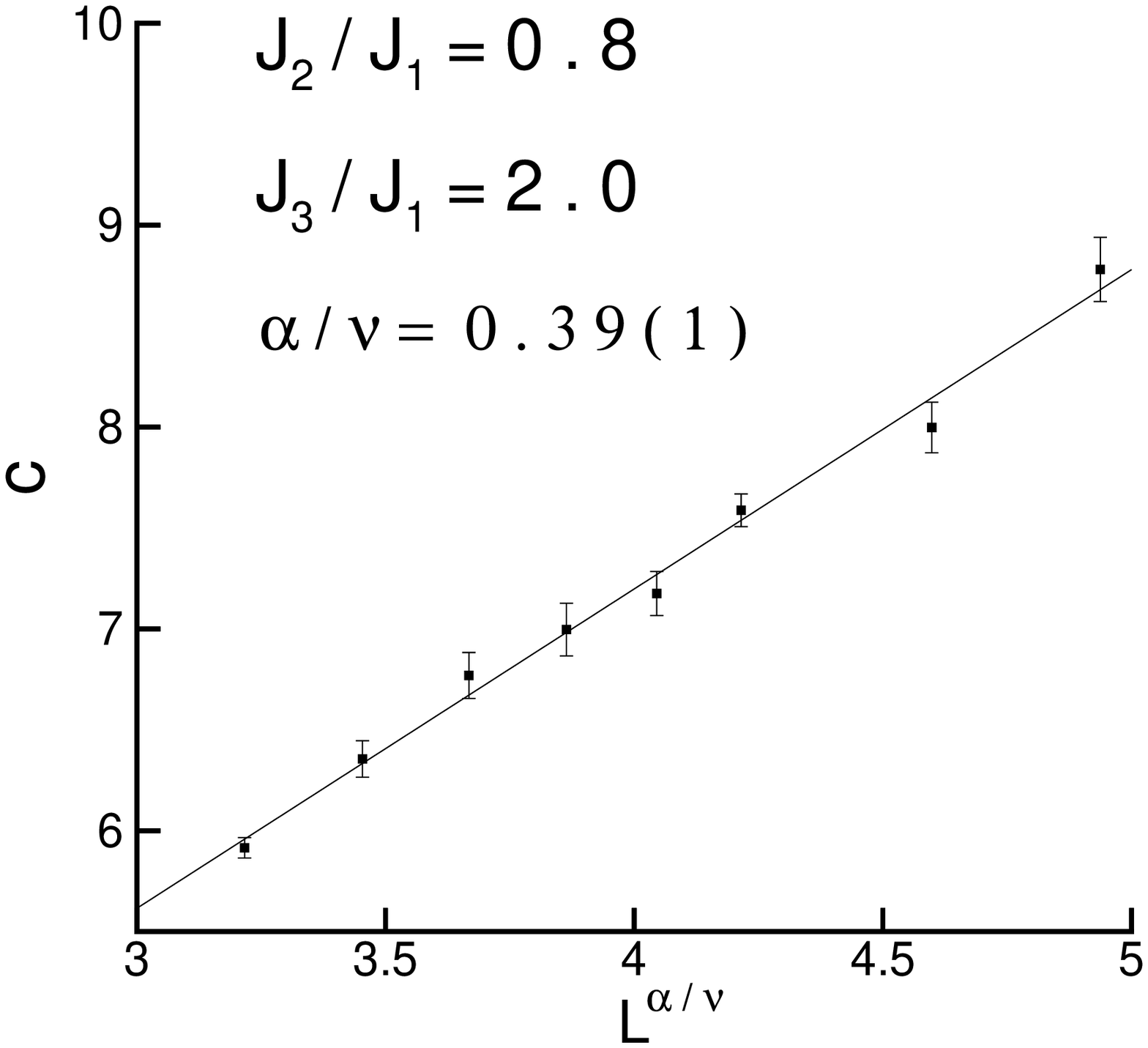} \narrowtext \caption{
Size dependence of specific-heat  at $T_{c}=1.257$ for
$J_{1}=1.0$, $J_{3}=2.0$ and $J_{2}=0.8$. The solid  line is
obtained by fitting the data to Eq(\ref{sh}).}
\end{figure}
%%%%%%%%%%%%%%%%%%%%%%%%%%%%%%%%%%%%%%%%%%%%%%%%%%%%%%%%%%%%%%%%%%%%%%%%%%%

\end{document}